\documentclass[showpacs]{revtex4}
\usepackage{hyperref}
\usepackage{amsfonts}
\usepackage{amsmath}
\usepackage{amssymb}
\usepackage{graphicx}

\begin{document}

\title{Reconstructing the interaction between the dark matter and
holographic dark energy}
\author{Shao-Feng Wu$^{1,2}$\footnote{%
Corresponding author. Email: sfwu@shu.edu.cn; Phone: +86-021-66136202.},
Peng-Ming Zhang$^{3,4}$, and Guo-Hong Yang$^{1,2}$}
\affiliation{$^{1}$Department of Physics, College of Science, Shanghai University,
Shanghai, 200444, P. R. China}
\affiliation{$^{2}$The Shanghai Key Lab of Astrophysics, Shanghai, P. R. China}
\affiliation{$^{3}$Center of Theoretical Nuclear Physics, National Laboratory of Heavy
Ion Accelerator, Lanzhou 730000, P. R. China}
\affiliation{$^{4}$Institute of Modern Physics, Lanzhou, 730000, P. R. China}

\begin{abstract}
We reconstruct the interaction rate between the dark matter and the
holographic dark energy with the parameterized equation of states and the
future event horizon as the infrared cut-off length. It is shown that the
observational constraints from the 192 SNIa and BAO measurement permit the
negative interaction in the wide region. Moreover, the usual
phenomenological descriptions can not describe the reconstructed interaction
well for many cases. The other possible interaction is also discussed.
\end{abstract}

\maketitle

\section{Introduction}

In the modern cosmology, a `dark energy' (DE) with negative pressure is
suggested to be responsible for the current acceleration of the universe.
The simplest candidate of DE is the cosmological constant, which does nicely
well at the pragmatic observational level, but entails the serious
theoretical difficultly: the cosmological constant problem and the
coincidence problem. Explanations of DE have been sought within a wide range
of physical phenomena, including some exotic fields, modified gravity
theories, and so on -see \cite{Padmanabhan} and references therein. Among
the most recent generic proposals, the model inspired by the holographic
idea \cite{Hooft}, that the quantum zero-point energy of a system cannot
exceed the mass of a black hole with same size, has been put forward to
explain the DE \cite{Cohen,Hsu,Li}. This DE density can be determined in
terms of the horizon radius of the universe, corresponding to relate the UV
cutoff of a system to its IR cutoff in the quantum field theory. There are
usually three choices for the horizon radius supposed to provide the IR
cutoff, with different degrees of success, namely the Hubble horizon, the
particle horizon, and the future event horizon. The event horizon may be
better, since in this case the DE can drive the present accelerated
expansion and the coincidence problem can be resolved by assuming an
appropriate number of e-folding of inflation \cite{Li,Li2,Lee}.

Most discussions on DE models rely on the fact that both dark matter (DM)
and DE only couple gravitationally. However, given their unknown nature and
the symmetry that would impose a vanishing interaction is still to be
discovered, an entirely independent behavior between dark sectors is very
special. Moreover, since DE must be accreted by massive compact objects like
black holes and neutron stars, in a cosmological context the energy transfer
from DE to DM may be small but must be non-vanishing. The interaction
hypothesis was first introduced by Wetterich \cite{Wetterich} to discuss the
cosmological constant problem in the light of dilatation symmetry and its
anomaly. Then cosmological consequences of a scalar field coupled to the
matter were studied in \cite{Wetterich1}. It was found that the coupling
quintessence models may give the scaling attractors providing an accelerated
expansion at the present time and alleviate the coincidence problem \cite%
{Amendola2}. The interaction also appears in the context of modified gravity
models \cite{Pavon3}. More possibility that DE and DM can interact has been
studied in \cite{Anderson,Mangano,Farrar,Elizalde,Setare,Chen1}. Confronted
to cosmological data, it was found that an appropriate interaction can
influence the perturbation dynamics, and the lowest multipoles of the CMB
spectrum \cite{Zimdahl,Wang1}, and could be inferred from the expansion
history of the Universe, as manifested in the supernova data together with
CMB and large-scale structure \cite{Guo,Feng,Amendola3}. In addition, it was
suggested that the dynamical equilibrium of collapsed structures would be
affected by the coupling of DE to DM \cite{Kesden,Bertolami}. The
interaction was first connected to holography by Horvat \cite{Horvat} who
argued that scaling of the cosmological constant stemming from the
zero-point energy in quantum field theory possibly implies a non-vanishing
coupling of the cosmological constant with DM. In the holographic DE model
with the Hubble horizon as the IR cutoff, the interaction can be available
to derive the present accelerated expansion and alleviate the coincidence
problem \cite{Pavon2}. In the interacting model with the event horizon as
the IR cutoff, it was shown that the equation of state (EoS) of DE can
accommodate the dynamically evolving behavior of crossing the phantom divide
\cite{Wang2}, which suggested by recent most observational probes \cite%
{WangY}.

Although the interaction is important in studying the physics of DE, it will
not be possible to derive the precise form of the interaction from first
principles unless the nature of both dark sectors was known. Usually, the
coupling is determined from phenomenological requirements \cite%
{Amendola2,Pavon2}. In view of the continuous equations of DE density $\rho
_{d}$ and DM density $\rho _{m}$, the coupling $Q$ must be a function of
densities multiplied by a quantity with units of inverse of time, which has
an obvious choice as Hubble time $H^{-1}$. Thus, one may write the coupling
as%
\begin{equation}
Q=Q(H\rho _{m},H\rho _{d}),  \label{Q}
\end{equation}%
which leads $Q\simeq \lambda _{m}H\rho _{m}+\lambda _{d}H\rho _{d}$ from the
first order terms in the power law expansion. Assuming that the ratio $%
r=\rho _{m}/\rho _{d}$ might be piecewise constant, the linear parameters
are usually set to $\lambda _{m}\simeq \lambda _{d}$ and even $\lambda
_{m}\simeq 0$ or $\lambda _{d}\simeq 0$ for simplicity. Considering the
couplings are terms in the Lagrangian which mix both DE and DM, one may
further suppose that they could be parameterized by some product of the
densities of DE and DM, such as the simplest $Q\simeq \lambda \rho _{m}\rho
_{d}$ \cite{Mangano}. Besides these phenomenological descriptions, various
proposals at the fundamental level have been tried to account for the
coupling, including the dependence of the matter field on the scalar field
\cite{Piazza} or expressing the cosmological constant as a function of the
trace of the energy-momentum tensor \cite{Poplawski}. Recently, an
interesting thermodynamical description of interaction between holographic
DE and DM has been proposed in \cite{Wang3}, where it was assumed that in
the absence of the coupling the DE and DM remain in separate thermal
equilibrium, then a small interaction can be viewed as a stable thermal
fluctuation that brings a logarithmic correction to the equilibrium entropy
of DE and DM. Other specific coupling which was assumed from the outset can
be found in \cite{Anderson,Amendola1,Das}.

The main aim of this work is to reconstruct the coupling using the recent DE
probes (the Baryon Acoustic Oscillation (BAO) measurement at $z=0.35$ from
the Sloan Digital Sky Survey \cite{Eisenstein} and the re-compiled 192 Type
Ia Supernovae (SnIa) samples \cite{Davis}, consisting 60 points from ESSENCE
(\textquotedblleft Equation of State: Supernovae trace Cosmic
Expansion\textquotedblright ) supernova survey \cite{Wood}, 57 points from
Supernovae Legacy Survey \cite{Astier}, 45 points nearby Supernovae \cite%
{Riess}, and 30 points detected by the Hubble Space Telescope \cite{Riess1}%
). We will focus on the holographic DE model and choose the future event
horizon as the IR cutoff.

The model with the Hubble horizon has been studied recently in \cite{Sen}
and it was found that the reconstructed interaction is always positive in 1$%
\sigma $ region. This seems to corroborate the recent argument that the
negative interaction violates the second law of thermodynamics which
inquires the energy transfer from DE to DM rather than otherwise \cite%
{Pavon1}. However, it should be noticed that there are some problems in the
thermodynamics of DE, such as the negative entropy \cite{Odintsov}, and the
generalized thermodynamical second law indeed breaks down in the universe
with phantom-dominated DE \cite{Izquierdo,Mohseni}. These results suggest
that one should consider the thermodynamical properties of DE with wide
possibilities. Hence, it is interesting to see whether the positive
interaction is a robust result for other models, such as the present model
with the IR cutoff as the future event horizon.

Considering the time varying DE gives a better fit than a cosmological
constant and in particular most of the observational probes indeed mildly
favor dynamical DE crossing the phantom divide at $z\sim 0.2$ \cite{WangY},
we will employ two commonly used parameterizations \cite%
{Jassal,Feng,Chevallier,Linder}, namely%
\begin{equation}
w_{A}=w_{0}+w_{1}(1-a)=w_{0}+w_{1}\frac{z}{1+z},  \label{w1}
\end{equation}%
which has been used in \cite{Sen}, and%
\begin{equation}
w_{B}=w_{0}+w_{1}(1-a)a=w_{0}+w_{1}\frac{z}{\left( 1+z\right) ^{2}}.
\label{w2}
\end{equation}%
It should be noticed that the different parameterizations are beneficial to
control some amount of parameterization dependence. After reconstructing the
interaction, we will further compare it with the usual phenomenological
models and the recent thermodynamical description.

\section{Reconstruction}

Let us begin with the Friedmann equations%
\begin{equation}
H^{2}=\frac{1}{3}\left( \rho _{m}+\rho _{d}\right) ,  \label{H1}
\end{equation}%
\begin{equation}
\dot{H}=-\frac{1}{2}(\rho _{m}+\rho _{d}+p_{m}+p_{d}),  \label{H2}
\end{equation}%
where we have normalized $8\pi G=1$ for conventions. The total energy
density $\rho =\rho _{m}+\rho _{d}$ satisfies a conservation law. However,
since we consider the interaction between DE and DM, $\rho _{m}$ and $\rho
_{d}$ do not satisfy independent conservation laws, they instead satisfy two
continuous equations%
\begin{equation}
\dot{\rho}_{m}+3H\rho _{m}=Q,  \label{rom}
\end{equation}%
\begin{equation}
\dot{\rho}_{d}+3H\left( 1+w_{d}\right) \rho _{d}=-Q,  \label{rod}
\end{equation}%
where $w_{d}$ is the EoS of DE, and $Q$ denotes the interaction term.
Without loss of generality, we will write the interaction as $Q=\rho
_{d}\Gamma $, where $\Gamma $ is an unknown function.

Using the the ratio of energy densities $r=\rho _{m}/\rho _{d}$, we have%
\begin{equation}
\frac{\dot{H}}{H^{2}}=-\frac{3}{2}\frac{1+w_{d}+r}{1+r}.  \label{Ht}
\end{equation}%
from Eq. (\ref{H2}). From Eq. (\ref{rom}), we have%
\begin{equation*}
\dot{r}=-r\frac{\dot{\rho}_{d}}{\rho _{d}}-3Hr+\Gamma ,
\end{equation*}%
which can be recast as
\begin{equation}
\dot{r}=(1+r)\Gamma +3Hrw_{d}  \label{rTao}
\end{equation}%
from Eq. (\ref{rod}). Eliminating the $\Gamma $ in above two equations, we
obtain%
\begin{equation}
\frac{-r\dot{r}-3Hrw_{d}}{1+r}=r\frac{\dot{\rho}_{d}}{\rho _{d}}+3Hr.
\label{rrot}
\end{equation}%
\

Until now, we have not specified the concrete DE density. We will focus on
the holographic DE model. Followed \cite{Li} by choosing the future event
horizon%
\begin{equation}
R_{E}=a\int_{a}^{\infty }\frac{dx}{Hx^{2}}  \label{RE}
\end{equation}%
as the IR cutoff, the holographic DE density is $\rho _{d}=3c^{2}R_{E}^{-2}$%
, where $c^{2}$ is a constant and the Planck mass has been taken as unit.
The most possible theoretical value of $c$ is one \cite{Li,Li2}, indicating
that the total energy from DE must be determined by the Schwarzschild
relation. Taking the derivative with respect to $t$, the evolution of the
horizon can be determined by%
\begin{equation*}
\dot{R}_{E}=HR_{E}-1.
\end{equation*}%
Defining $\Omega _{d}=\rho _{d}/3H^{2}$, we have $\Omega _{d}=1/(1+r)$ and $%
R_{E}=c\sqrt{1+r}/H$. Thus, Eq. (\ref{rrot}) can be recast as%
\begin{equation}
\frac{-r\dot{r}-3Hrw_{d}}{1+r}=r\frac{-2H\left( c\sqrt{1+r}-1\right) }{c%
\sqrt{1+r}}+3Hr.  \label{rt}
\end{equation}%
Obviously, $r$ is not a constant in general. This is different with the case
in \cite{Sen} where $r$ is a constant since the IR cutoff was chosen as
being the Hubble scale. Replacing the time $t$ as the redshift $z=\left(
1-a\right) /a$, we can rewrite Eq. (\ref{rt}) as%
\begin{equation}
\frac{-(1+z)rr^{\prime }+3rw_{d}}{1+r}=r\frac{2\left( c\sqrt{1+r}-1\right) }{%
c\sqrt{1+r}}-3r,  \label{rz}
\end{equation}%
where the prime denotes the derivative with respect to $z$. Similarly, Eq. (%
\ref{Ht}) reads%
\begin{equation}
\frac{H^{\prime }}{H}=\frac{3}{2(1+z)}\frac{1+w_{d}+r}{1+r}.  \label{Hz}
\end{equation}%
It is interesting to find that Eqs. (\ref{rz}) and (\ref{Hz}) determine the
evolvement of $r$ and $H$, if we know the EoS $w_{d}$. In the normal
interacting DE model, one assumes the explicit interaction form to give out $%
w_{d}$. For our aim, we will use the two commonly used parameterizations of $%
w_{d}$ to determine the dynamics of our model, and reconstruct the
interaction rate $\Gamma /3H$%
\begin{equation*}
\frac{\Gamma }{3H}=-\frac{(1+z)r^{\prime }+3rw_{d}}{3\left( 1+r\right) }
\end{equation*}%
from the recent observational datasets.

We will use the re-compiled 192 SnIa samples ($0<z<1.8$) combined with the
recent BAO measurement from SDSS to reconstruct the interaction rate. There
are other DE observational probes, including the three-year WMAP CMB shift
parameter, the X-ray gas mass fraction in clusters, the linear growth rate
of perturbations at $z=0.15$ as obtained from the 2dF galaxy redshift
survey, and the look back age data. However, since the parameterizations of $%
w_{d}$ Eqs. (\ref{w1}) and (\ref{w2}) are motivated to accommodate the
dynamically evolving behavior of crossing the phantom divide at recent
epoch, and our model has not included the radiation and the baryonic matter
which may be important in the early, we will not use the WMAP CMB shift
parameter which focuses on the high redshift region. Besides, for
simplicity, we do not adopt other probes of DE which have large relative
errors compared with SnIa, CMB, and BAO probes \cite{Nesseris}.

\begin{figure}[tbp]
\begin{center}
\includegraphics[
height=2in, width=6in ]{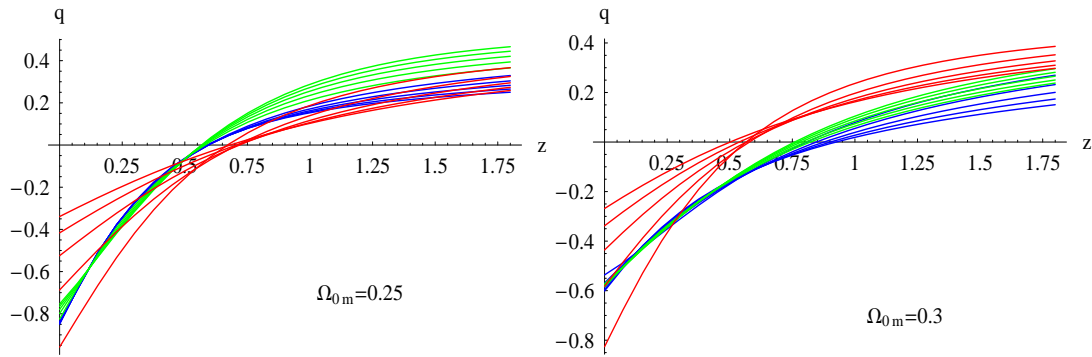}
\end{center}
\caption{The deceleration parameter $q$ with the best-fit values of equation
of states $w_{d}$. The red, green, and blue lines denote the models without
interaction, with interaction and $w_{d}=w_{A}$, and with interaction and $%
w_{d}=w_{B}$, respectively. For each model, there are five lines from top to
down at early taking different $c$ from 0.6 to 1.4. }
\label{q}
\end{figure}

\begin{figure}[tbp]
\begin{center}
\includegraphics[
height=1.5in, width=7in ]{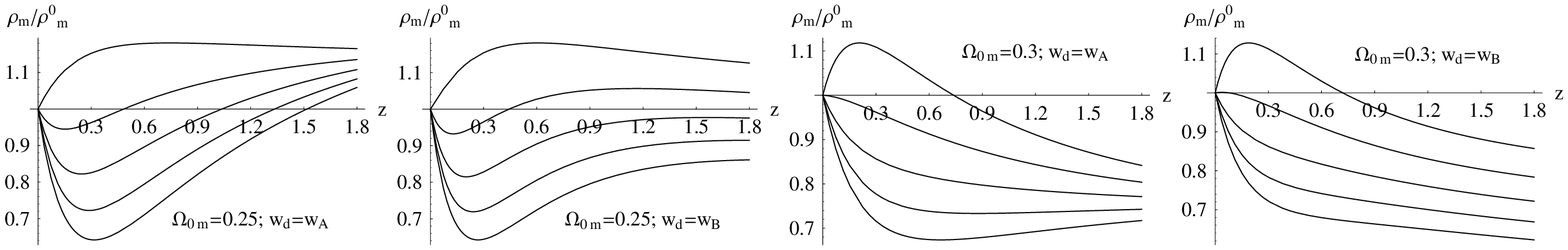}
\end{center}
\caption{The ratios $\protect\rho _{m}/\protect\rho _{m}^{0}$ for different $%
\Omega _{0m}$, $w_{d}$, and $c$. In each panel, $c$ increases from top to
bottom.}
\label{roi}
\end{figure}

\begin{figure}[tbp]
\begin{center}
\includegraphics[
height=1.5in, width=7in ]{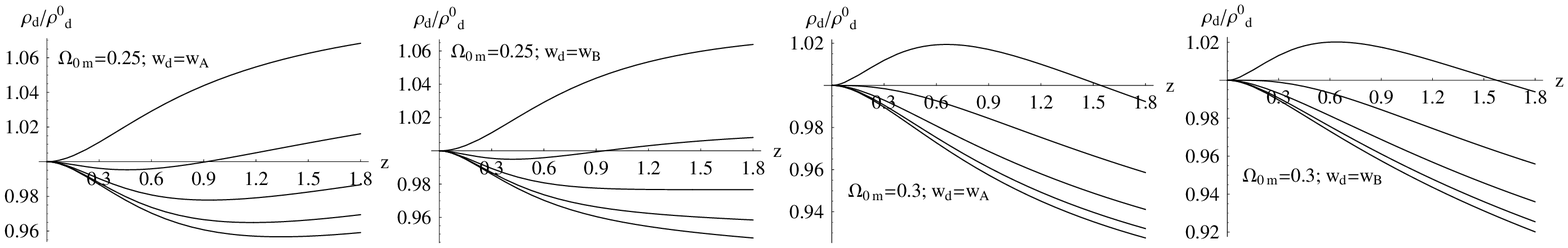}
\end{center}
\caption{The ratios $\protect\rho _{d}/\protect\rho _{d}^{0}$ for different $%
\Omega _{0m}$, $w_{d}$, and $c$. In each panel, $c$ increases from top to
bottom.}
\label{roid}
\end{figure}

As usually, we will fix DM density parameter as $\Omega _{0m}=0.3$\ or $0.25$%
\ to include the best-fit value of $\Omega _{0m}=0.27$\ from 5-year WMAP
data. In general, they are sufficiently representative. Moreover, it is
convenient to compare our reconstructed interaction to the interaction
reconstructed in \cite{Sen} where these two DM density parameters are used.
We will consider five indicative different values of the constant $c$ near
one. The best-fit values with 1$\sigma $ error bars for the parameters $%
w_{0} $ and $w_{1}$ are given in TABLE \ref{tab1}, \ref{tab2}.
\begin{table}[th]
\begin{tabular}{|c|c|c|c|c|c|c|}
\hline
& $c$ & $0.6$ & $0.8$ & $1.0$ & $1.2$ & $1.4$ \\ \hline
$w_{A}$ & $w_{0}$ & -1.19$\pm $0.18 & -1.16$\pm $0.16 & -1.14$\pm $0.16 &
-1.13$\pm $0.15 & -1.12$\pm $0.14 \\ \cline{2-7}
& $w_{1}$ & 0.60$\pm $1.26 & 1.09$\pm $1.07 & 1.35$\pm $0.97 & 1.50$\pm $0.90
& 1.60$\pm $0.85 \\ \hline
$w_{B}$ & $w_{0}$ & -1.21$\pm $0.24 & -1.21$\pm $0.22 & -1.21$\pm $0.20 &
-1.21$\pm $0.19 & -1.20$\pm $0.19 \\ \cline{2-7}
& $w_{1}$ & 0.99$\pm $2.11 & 1.83$\pm $1.83 & 2.29$\pm $1.66 & 2.58$\pm $1.55
& 2.78$\pm $1.47 \\ \hline
\end{tabular}%
\caption{The best fit values with 1$\protect\sigma $ error bars for the
parameters of $w_{d}$ with DM density parameter $\Omega _{0m}$ $=0.25$.}
\label{tab1}
\end{table}
\begin{table}[th]
\begin{tabular}{|c|c|c|c|c|c|c|}
\hline
& $c$ & $0.6$ & $0.8$ & $1.0$ & $1.2$ & $1.4$ \\ \hline
$w_{A}$ & $w_{0}$ & -1.03$\pm $0.20 & -1.03$\pm $0.18 & -1.03$\pm $0.16 &
-1.03$\pm $0.16 & -1.02$\pm $0.15 \\ \cline{2-7}
& $w_{1}$ & 0.48$\pm $1.28 & 0.48$\pm $1.09 & 0.48$\pm $0.98 & 0.70$\pm $0.92
& 0.85$\pm $0.87 \\ \hline
$w_{B}$ & $w_{0}$ & -0.99$\pm $0.26 & -1.02$\pm $0.23 & -1.04$\pm $0.22 &
-1.05$\pm $0.21 & -1.05$\pm $0.20 \\ \cline{2-7}
& $w_{1}$ & -1.14$\pm $2.20 & 0.03$\pm $1.90 & 0.68$\pm $1.73 & 1.09$\pm $%
1.62 & 1.37$\pm $1.54 \\ \hline
\end{tabular}%
\caption{The best fit values with 1$\protect\sigma $ error bars for the
parameters of $w_{d}$ with DM density parameter $\Omega _{0m}$ $=0.3$.}
\label{tab2}
\end{table}
One can find that the phantom divide crossing in recent epoch is always
permitted in 1$\sigma $ region. The deceleration parameter with these
best-fit equation of states is plotted in FIG. \ref{q},
where the acceleration (super-acceleration, at most cases) in recent epoch
is achieved. It is interesting to see that current deceleration parameter in
interacting model is almost not affected by the DE parameter $c$, according
to the almost same parameters $w_{0}$ in TABLE \ref{tab1}, \ref{tab2}. We
also show the evolution of $\rho _{m}$ compared its non-interacting version $%
\rho _{m}^{0}$ in FIG. \ref{roi}.
The ratios change strongly in recent epoch which implies the interaction
plays important role in that time. The correspnding plots for the dark
energy component are given in FIG. \ref{roid}. Since the DE density is
always positive, this suggests that the interaction can be negative but can
not be too negative when $w_{d}>-1$, which is possible at least in the past.

\begin{figure}[tbp]
\begin{center}
\includegraphics[
height=3.5in, width=7in ]{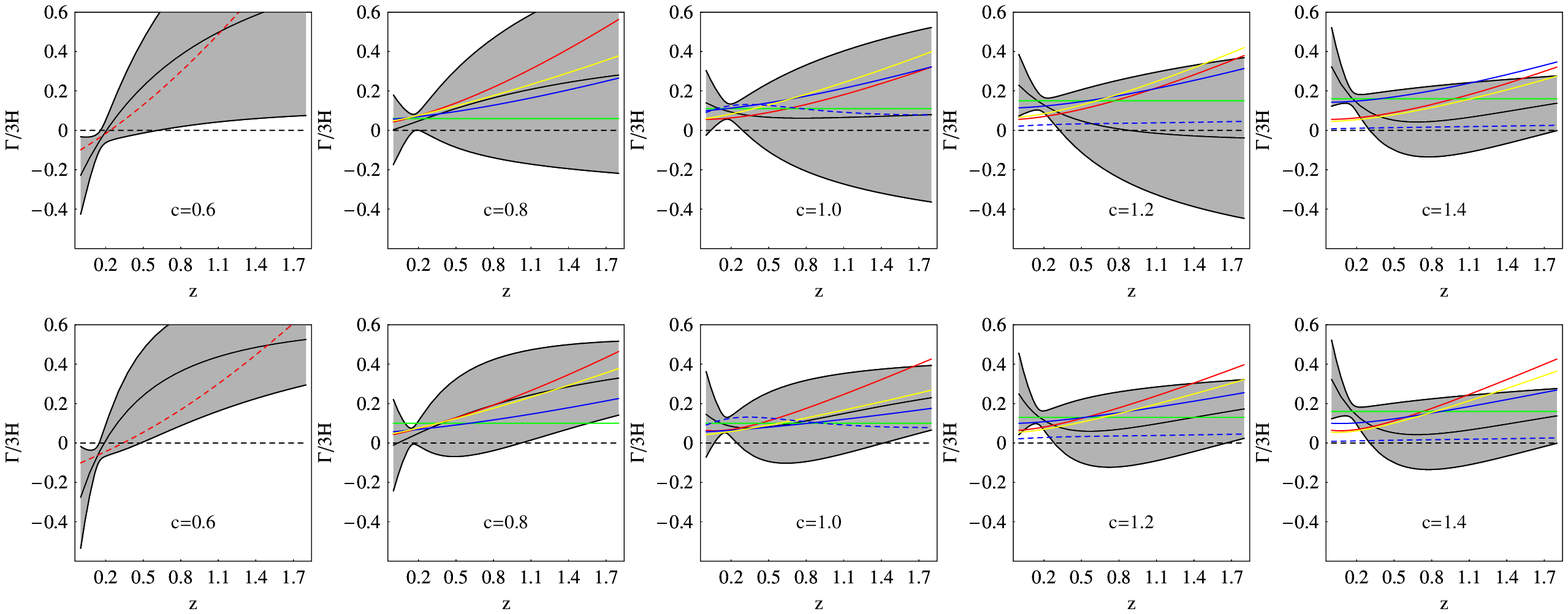}
\end{center}
\caption{Interaction rates for $\Omega _{0m}$ $=0.3$ with respect to
redshift $z$. Above and below panels denote $w_{d}=w_{A}$ and $w_{B}$,
respectively. The black curves and the grey region between them indicate the
best fit curve of the reconstructed interaction rate $\Gamma /3H$ and the
region in 1$\protect\sigma $ confidence level. The black dashing lines
indicate $\Gamma /3H=0$. The red, green, blue, and yellow curves indicate
the $R_{i}$ from $i=1$ to $i=4$ with four constants $\protect\lambda $. For
example, we give $\protect\lambda =0.13,0.11,0.07,0.2$ in turn for $%
w_{d}=w_{A}$ and $c=1$. The blue dashing lines indicative the interaction
rate $R_{5}$. The red dashing lines in the case with $c=0.6$ indicative the
interaction rate $R_{6}$ with two nonzero constants. For the above panel,
they are $\protect\lambda _{1}=0.17,\protect\lambda _{2}=-0.16$; For the
below panel, they are $\protect\lambda _{1}=0.11,\protect\lambda _{2}=-0.15.$%
}
\label{int1}
\end{figure}

\begin{figure}[tbp]
\begin{center}
\includegraphics[
height=3.5in, width=7in ]{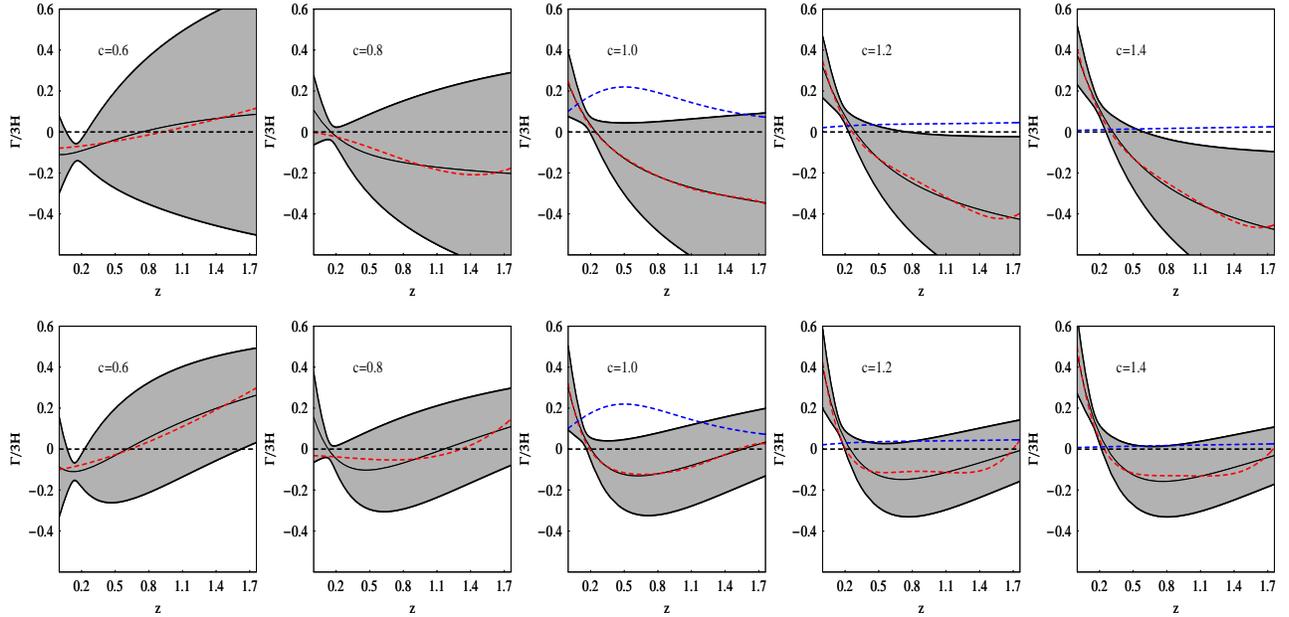}
\end{center}
\caption{Interaction rates for $\Omega _{0m}$ $=0.25$. The blue dashing
lines indicative the interaction rate $R_{5}$. The red dashing lines
indicative the interaction rate $R_{6}$ with some constants $\protect\lambda %
_{i}$ ($i=1,\cdots ,5$).}
\label{int2}
\end{figure}

With these best-fit values, we reconstruct the interaction rates, see FIG. %
\ref{int1} and FIG. \ref{int2}
This is one of the main result of this paper. It should be stressed that the
negative interaction is permitted in the wide region of FIG. \ref{int1} and
FIG. \ref{int2}. This is contrast to the result in \cite{Sen} where the
reconstructed interaction is always positive, and the argument that the
second law of thermodynamics which imposes energy transfer from DE to DM
\cite{Pavon1}. If the second law is true, the permitted region of the
interaction rates must be reduced, as done in \cite{Karwan}. However, for
generality and considering the thermodynamical properties of DE is not
clear, we will not restrict the interaction to be positive.

Moreover, the effective EoS $w_{eff}=w_{d}+\frac{\Gamma }{3H}$\ can be
obtained since we have known $w_{d}$\ and reconstructed the $\Gamma /3H$.
From TABLE \ref{tab1}, \ref{tab2}, and FIG. \ref{int1}, \ref{int2}, one can
know that the effective EoS may be bigger or smaller than $-1$\ for
different parameters $\Omega _{0m}$\ and $c$. If one requires an effective
phantom-like DE, the presence of a coupling makes the parameters have more
possible values than the case of absence of a coupling, where the EoS is
determined by $w_{d}=-\frac{1}{3}-\frac{2}{3c}\sqrt{\Omega _{d}}$\ which
imposes $c<\sqrt{\Omega _{d}}$\ for $w_{d}<-1$. Moreover, it is possible to
have effective EoS $w_{eff}$\ smaller than $-1$\ but $w_{d}$\ bigger than $-1
$\ when the interaction is negative. This suggests that the negative
interaction is interesting because DE with $w_{d}>-1$\ is easily accepted
(The scalar field with $w_{d}<-1$\ will break the zero energy condition.)
and the effective EoS of DE $w_{eff}<-1$\ can fit the cosmological data
better.

\section{Comparison}

In the following, we will compare the reconstructed interaction rate with
other descriptions. Let us begin with the usual phenomenological
descriptions. Usually there are four different choices of $Q_{i}$ ($%
i=1,2,3,4 $), which can be expressed as $3\lambda H\rho _{d}$, $3\lambda
H\rho _{m}$, $3\lambda H\left( \rho _{m}+\rho _{d}\right) $, and $\lambda
\rho _{m}\rho _{d}$, respectively. To compare them with the reconstructed
interaction rate $\Gamma /3H$, we define four interaction rates $%
R_{i}=Q_{i}/\left( 3H\rho _{d}\right) $, which can be determined by $r$ and $%
H$ (which have been solved from Eqs. (\ref{rz}) and (\ref{Hz})), namely $%
R_{1}=\lambda $, $R_{2}=\lambda r$, $R_{3}=\lambda (1+r)$, and $%
R_{4}=\lambda Hr/(1+r)$. Observing FIG. \ref{int1}, the indicative curves of
the usual interaction rates $R_{i}$ ($i=1,2,3,4$) show that in many cases
they are not favored in 1$\sigma $ region. For example, see the case of $%
c=1.2$, where the interaction rates $R_{2}$ (green line) and $R_{3}$ (blue
line) are permitted in 1$\sigma $ region, but the other interaction rates $%
R_{1}$ (red line) and $R_{4}$ (yellow line) are not permitted whether $%
\lambda $ increases or decreases. For the case $c=0.6$, the reconstructed
interaction rate changes at recent epoch from positive to negative. This
tendency is more general in FIG. \ref{int2} with small DM parameter $\Omega
_{0m}$ $=0.25$. Thus, one may suspect that the four phenomenological
descriptions of the interaction which have the definitive sign may be not
suitable.

We will also consider the recent thermodynamical description of the
interaction. Defining $Q_{5}=3Hb^{2}(\rho _{m}+\rho _{d})$, the
thermodynamical interaction rate $b^{2}$ can be determined by following
equations (see \cite{Wang3} in detail):%
\begin{equation*}
b^{2}=\frac{2\Omega _{d}^{3/2}}{3c}\left[ -1+\frac{H^{2}\sqrt{\Omega _{d}}}{%
\left( H^{0}\right) ^{2}\sqrt{\Omega _{d}^{0}}}\frac{\sqrt{\Omega _{d}^{0}}%
/c-1}{\sqrt{\Omega _{d}}/c-1}\right] +\frac{1}{12\pi c^{2}}\frac{H^{2}}{c/%
\sqrt{\Omega _{d}}\left( c/\sqrt{\Omega _{d}}-1\right) }\frac{\sqrt{\Omega
_{d}}}{\sqrt{\Omega _{d}^{0}}}\left( -\Omega _{d}^{0}\right) ^{\prime }(1+z),
\end{equation*}%
\begin{equation*}
-(1+z)\frac{\Omega _{d}^{\prime }}{\Omega _{d}}+(\Omega _{d}-1)+\frac{2\sqrt{%
\Omega _{d}}}{c}(\Omega _{d}-1)=-3b^{2},
\end{equation*}%
\begin{equation*}
\frac{-(1+z)H^{\prime }}{H}=\frac{\sqrt{\Omega _{d}}}{c}-1+\frac{(1+z)\Omega
_{d}^{\prime }}{2\Omega _{d}},
\end{equation*}%
where the zero superscript of $\Omega _{d}^{0}$ and $H^{0}$ indicates
absence of interaction. We can relate $b^{2}$ to a new interaction rate%
\begin{equation*}
R_{5}=\frac{Q_{5}}{3H\rho _{d}}=\frac{b^{2}(\rho _{m}+\rho _{d})}{\rho _{d}}=%
\frac{b^{2}}{\Omega _{d}}.
\end{equation*}%
In FIG. \ref{int1} and FIG. \ref{int2} (blue dashing lines), one can find
that $R_{5}$ is almost not favored in 1$\sigma $ region (Note we have not
shown the cases with $c<1$ where $b^{2}$ has not the real solution and the
parameterization of $w_{d}$ is not needed to determine $R_{5}$ since the EoS
in this model is determined by the $b^{2}$.).

It is interesting to ask whether the reconstructed interaction can be
described well by a more general interaction form. A natural candidate is to
expand the phenomenological description (\ref{Q}) up to the second order,
namely%
\begin{equation*}
Q_{6}=3\lambda _{1}H\rho _{m}+3\lambda _{2}H\rho _{d}+\lambda _{3}H^{2}\rho
_{m}\rho _{d}+\lambda _{4}H^{2}\rho _{m}^{2}+\lambda _{5}H^{2}\rho _{d}^{2},
\end{equation*}%
with the interaction rate%
\begin{equation*}
R_{6}=\frac{Q_{6}}{3H\rho _{d}}=\lambda _{1}r+\lambda _{2}+\lambda _{3}\frac{%
H^{3}r}{1+r}+\lambda _{4}\frac{H^{3}r^{2}}{1+r}+\lambda _{5}\frac{H^{3}}{1+r}%
.
\end{equation*}%
We find that it indeed works well, see the red dashing line in FIG. \ref%
{int1} and FIG. \ref{int2}. However, this parameterization only has
theoretical interest since it contains too many parameters. Moreover,
numerical calculations prove that this parameterization can not be reduced
to include only the two first order terms, which has been studied recently
in \cite{Karwan}, if we need it being permitted in 1$\sigma $ region for all
cases.

\section{Summary}

We have reconstructed the interaction term between the holographic DE and
DM, using the re-compiled 192 SnIa samples combined with the recent BAO
measurement. The DE parameter $c$ is assumed near one. The two common used
parameterizations of $w_{d}$ are considered in 1$\sigma $ region. It is
found that the present accelerated expansion of universe is achieved and the
phantom behavior of DE is permitted. We illustrate that the negative
interaction is permitted in the wide region. Hence, contrast to the DE model
studied in \cite{Sen}, where the reconstructed interaction is always
positive in 1$\sigma $ region, we can not obtain the favor from the DM
observation for the recent argued thermodynamical second law which imposes
the energy transfer from DE to DM. This suggest us to keep wide
possibilities of the thermodynamical properties of DE.

We show that the four usual phenomenological descriptions can not describe
the reconstructed interaction well for many cases. Specially for the case
with small DM parameter $\Omega _{0m}$ $=0.25$, the interaction rate has a
trend to change sign at recent epoch. This is at variance with the usual
phenomenological interacting terms which have the definitive sign. We
further illustrate that the recent thermodynamical description of the
interaction is not favored. Our work stimulates one to seek a more suitable
interaction.

It would be interesting to confront our model to more observations, such as
CMB angular power and large scale structure, which will further constrain
the interaction rate and make clear if the negative interaction and phantom
DE are still to be permitted. Another interesting work is to consider the
perturbation evolution in our model. Recently, it has been found that some
types of interaction will lead to instability under the curvature
perturbation \cite{Valiviita,hjh}. It may provide some restrictions on the
form of interaction. We will work on these directions in the future.

\begin{acknowledgments}
This work was supported by NSFC under Grant Nos. 10847102 and 10604024, the
Shanghai Research Foundation No. 07dz22020, the CAS Knowledge Innovation
Project Nos. KJcx.syw.N2, the Shanghai Education Development Foundation, the
Innovation Foundation of Shanghai University, and Shanghai Leading Academic
Discipline Project S30105.
\end{acknowledgments}

\end{document}